\documentclass[a4paper,11pt]{article}

\usepackage{amsmath,amssymb}
\usepackage[latin1]{inputenc}
\usepackage{theorem}
\usepackage{latexsym}
\usepackage{amsfonts}
\usepackage{eufrak}
\usepackage{syntonly}
\usepackage[mathcal]{euscript}
\usepackage{pb-diagram}

\def\pb{\bar\partial}

\def\({\left(}
\def\){\right)}
\def\<{\left\langle\,}
\def\>{\, \right\rangle}
\def\[{\left[}
\def\]{\right]}
\def\d{\text{d}}

\def\a{\alpha}
\def\b{\beta}
\def\g{\gamma}
\def\l{\lambda}
\def\o{\omega}
\def\ob{\bar\omega}
\def\lb{\bar\lambda}
\def\ga{\gamma^}
\def\de{\delta}
\def\r{\rho}

\def\t{\theta}
\def\k{\kappa}
\def\p{\partial}
\def\ap{\alpha^{\prime}}
\def\deg{\rm {degree}}

\numberwithin{equation}{section}

\begin{document}

    \vskip1.1truein
    \centerline{ }
    \vskip 2cm
    
    \centerline{\Large{\bf One-loop Superstring Amplitude From}}
    \vskip 0.27cm
    \centerline{\Large{\bf Integrals on Pure
     Spinors Space}}
    \vskip 1.3cm
     \centerline{ Humberto Gomez
    \footnote{\tt 
    humgomzu@ift.unesp.br
    }
     }
    \vskip 25pt
    \centerline{\sl 
Instituto de F\'isica Te\'orica
UNESP - Universidade Estadual Paulista}
\centerline{\sl Caixa Postal 70532-2
01156-970 S\~ao Paulo, SP, Brazil}
     \vskip 2.5cm
    
\begin{abstract}
In the Type II superstring the 4-point function for massless NS-NS bosons at one-loop is well
known
\cite{tanii}\cite{dhoker phong s duality}. The overall constant factor in this amplitude is very
important because it needs to satisfy the unitarity and S-duality
conditions \cite{dhoker phong s duality}. This coefficient has not been computed in the pure spinor
formalism due to the
difficulty to solve the integrals on the pure spinors space. In this paper we compute it by using
the
non-minimal pure spinor formalism and we will show that the answer is in perfect agreement with the
one given in \cite{dhoker
phong
s duality}.

\end{abstract}                 
    %
      \vskip 2cm
    %
    %
    %
    %
    %
    
       \newpage


\tableofcontents

    \section{Introduction}

The pure spinor formalism has many advantages for computing scattering amplitudes compared to
the
RNS and
the GS formalism. For example, it does not have to deal with  worldsheet spin
structures \cite{nathan covariant superstring}\cite{dhoker phong
perturbation}, it has manifest Super-Poincar\'e invariance  and  incorporate in a natural way the
Ramond sectors. Nevertheless the formalism presents some difficulties, for example, the
normalization of
the integration measure in the pure spinors space, the computational difficulty to solve the
integrals in this space and 
the $S$ matrix unitarity has not been demostrated yet.

In this paper we will compute the one-loop scattering amplitude in the non-minimal pure spinor
formalism for
Type II superstrings and we will show that the overall constant factor is the same as the one given
in \cite{dhoker phong s duality}. Let's remember that this factor was also computed from the
unitarity condition \cite{tanii}. So, showing that the non-minimal pure spinor formalism predicts
the
same result as the RNS formalism  is a direct test of unitarity.\\ 
To compute the scattering amplitude we normalize the integration 
measure of the pure spinors space in the same way as the phase space in quantum mechanics is
normalized in the path integral, this is  because the pure spinor formalism is a first order
formalism.\\
To compute the integral on pure spinors space we use some tools of
algebraic geometry. We also show that this normalization in the amplitude does not require computing
functional determinants at all. This implies that computations using pure spinor formalism 
are
easier than the ones done in RNS or GS formalism.                 

This paper is organized as follows. In Section 2, the non-minimal pure spinor formalism
will be reviewed and the space time units will be defined. We will normalize the massless vertex
operator of the pure spinors formalism  to  coincide with the RNS normalization.
In Section 3, the 4-point one-loop scattering amplitude will be computed in the NS-NS sector
using the non-minimal pure spinor formalism, up to an integration on pure spinors space. In the
Subsection
3.1 we will give a review to the $x^{m}(z,\bar z)$ fields
contribution and we justify the normalization of the path integral measures. In the Subsection 3.2
we compute the contribution of the others fields and discuss biefly the modular
invariance of the scattering amplitude. We use some results found in \cite{nathan
mafra nonminimal}\cite{mafra 5 points}\cite{mafra 1 loop}\cite{mafra tesis} in which the authors
showed: 1) the equivalence between the kinematic factor of the non-minimal pure spinor formalism
and
the minimal pure spinors formalism, 2) the equivalence between the kinematic factor of the
minimal pure spinors formalism and the RNS formalism. At the end of the Section we find all the
factors in the 1-loop scattering amplitude, up to an integration over pure
spinors space.
In the last Section, we will compute the integral on the pure spinors space. This is the
most important Section of the paper and we suggest the reader check the Appendix beforehand, in
which we
apply the
tools
used to compute the integral in the pure spinors space in lower dimensions ($D=2n<10$). The aim is
to be more familiar with the concepts of algebraic
geometry involved in the computation.
In this Section we arrive to the following result
\begin{center}
\begin{tabular}{| c |}
 \cline{1-1}   \\
$\int_{\mathcal{O}(-1)}[\d\l]\wedge[\d\lb]\,e^{-a\,\l\lb}= (2\pi)^{11}(a^{8}\cdot
12\cdot 5)^{-1}\,,\qquad
a\in\mathbb{R}^{+}$\\
 \\
\cline{1-1}
\end{tabular}
\end{center}
where $\mathcal{O}(-1)$ is the line bundle blow-up at the origin with base space
$SO(10)/U(5)$. In others words, $\mathcal{O}(-1)$ is  the pure spinors space. Finally, with this
result we find the
overall constant factor, which is called $C_{1}$ \cite{dhoker phong s duality}. 

Our future goal is to
compute the overall constants factors at tree level, which we call $C_{0}$, and at two loops, called
$C_{2}$, in the
non-minimal pure spinor formalism \cite{work in progress} and
to show that the S-duality constraint ($C_{1}^{2}=2\pi^{2}C_{0}C_{2}$)\cite{dhoker phong s duality}
is a consequence of the identities for massless four-point kinematic factors \cite{mafra
identities}.

\section{Review on the non-minimal pure spinor formalism}

We will give a brief review of the non-minimal pure spinor formalism. The idea is to introduce
our own conventions and to normalize the massless vertex operator in the same way as in the D'Hoker,
Phong
and
Gutperle's paper \cite{dhoker phong s duality}.

The superstring theory action in the right sector of the non-minimal  pure spinor formalism
\cite{nathan
topological} is
given
by

\begin{equation}
S=\frac{1}{2\pi\alpha^{\prime}}\int_{\Sigma_{g}}\d^{2}z\,\,\(\p
x^{m}\bar\p
x_{m}+\alpha^{\prime}p_{\a}\bar\p\theta^{\a}-\alpha^{\prime}\omega_{\a}\bar\p\lambda^{\a}-
\alpha^{\prime}\bar\omega^{\a}\bar\p\bar\lambda_{\a}+\alpha^{\prime}s^{\a}\bar\p
r_{\a}\)
\end{equation}
where we define the space time dimensions of the variables and coupling constant $\ap$
as follows
\begin{eqnarray}
&&\[x^{m}\]=1,\quad \[\a^{\prime}\]=2,\quad
\[p_{\a}\]=\[\omega_{\a}\]=\[\bar\lambda_{\a}\]=\[r_{\a}\]=-1/2,\\
&&\[\theta^{\a}\]=\[\lambda^{\a}\]=\[\bar\omega^{\a}\]=\[s^{\a}\]=1/2.
\end{eqnarray}
The OPE's for the matter variables are easily computed
\begin{eqnarray}
x^{m}(z)x_{n}(w)\sim
-\frac{\a^{\prime}}{2}\delta^{m}_{n}\text{ln}|z-w|^{2},\quad
p_{\a}(z)\theta^{\beta}(w)\sim\frac{\delta^{\beta}_{\a}}{z-w}.
\end{eqnarray}
\\
The complex bosonic spinors  $\lambda^{\a}$ and $\bar\lambda_{\a}$
satisfy\footnote{the $\lb_{\a}$ spinor is treated as the complex conjugate of the $\l^{\a}$ spinor.}
the pure
spinor constraint
\begin{equation}
\lambda\gamma^{m}\lambda=\bar\lambda\gamma^{m}\bar\lambda=0,\qquad m=0,\,1,\,2,...,9
\end{equation}
and the fermionic spinor $r_{\a}$ satisfies the constraint
\begin{equation}
\bar\lambda\gamma^{m}r=0.
\end{equation}
Because of the constraints on $\l^{\a},\,\bar\l_{\a}$ and $r_{\a}$, their conjugate momenta
$\o_{\a},\,\bar\o^{\a}$ and $s^{\a}$ are defined up to a gauge tranformation, 
\begin{eqnarray}
&&\de\o_{\a}=\Lambda_{m}(\g^{m}\l)_{\a}\\
&&\de\bar\o^{\a}=\bar\Lambda_{m}(\g^{m}\bar\l)^{\a}-\phi_{m}(\g^{m}r)^{\a},\quad \de
s^{\a}=\phi_{m}(\g^{m}\bar\l)^{\a}\,\, ,
\end{eqnarray}
for arbitrary $\Lambda_{m},\,\bar\Lambda_{m}$ and $\phi_{m}$.

In the $U(5)$ variables the pure spinor constraints takes the following form \cite{nathan
topological}
\begin{eqnarray}
&&2\l^{+}\l^{a}-\frac{1}{4}\epsilon^{abcde}\l_{bc}\l_{de}=0,\qquad a,\,b,\,c,\,d,\,e=1,\,2,...,5\\
&& 2\l^{b}\l_{ab}=0.
\end{eqnarray}                            
where just five equations are linearly independent. In the chart $U_{+++++}=\{\l^{+}\neq 0\}$
these equations are solved by \cite{cartan}
\begin{eqnarray}\label{para}
\l^{+}=\g,\quad \l_{ab}=\g u_{ab},\quad \l^{a}=\frac{1}{8}\g\epsilon^{abcde}u_{bc}u_{de}.
\end{eqnarray}
As the $u_{ab}$ variables parametrize the projective pure spinors space, then it is clear that the
pure
spinors space is the total space of the $\mathcal{O}(-1)$ bundle over the projective pure spinors
space with blow-up at the origin ($\g=0$) \cite{nikita beta gamma}
\cite{harris} \cite{vafa}. 

In this chart, we can take the gauge $\o_{a}=\bar\o^{a}=0$ and the parametrization
\begin{eqnarray}
&&\o_{+}=\b-\frac{1}{2\g}v^{ab}u_{ab},\quad \o^{ab}=\frac{v^{ab}}{\g}, \\
&& \bar\o^{+}=\bar\b-\frac{1}{2\bar\g}\bar v_{ab}\bar
u^{ab},    \quad  \bar\o_{ab}=\frac{\bar v_{ab}}{\bar\g},
\end{eqnarray}
so the pure spinors action takes the form
\begin{equation}
S_{PS}=\frac{1}{2\pi}\int\d^{2}z\,\(\b\pb\g+\frac{1}{2}v_{ab}\pb u^{ab}
+\bar\b\pb\bar\g+\frac{1}{2}\bar v^{ab}\pb \bar u_{ab}\).
\end{equation}
With this action it is easy to get the OPE's
\begin{eqnarray}
&&\b(z)\g(w)\rightarrow (z-w)^{-1},\quad v^{ab}(z)u_{cd}(w)\rightarrow
\de^{a}_{[c}\de^{b}_{d]}(z-w)^{-1},\\
&&\bar\b(z)\bar\g(w)\rightarrow (z-w)^{-1},\quad\bar v_{ab}(z)\bar u^{cd}(w)\rightarrow
\de^{c}_{[a}\de^{d}_{b]}(z-w)^{-1}.
\end{eqnarray}
For the $s^{\a},\,r_{\a}$ fields the procedure is similar.


From the previous definitions of the space-time dimensions of the
fields and their OPEs we can get the following OPE's \cite{nathan multiloop}
\begin{eqnarray*}
&&d_{\a}=p_{\a}-\frac{1}{\ap}\g^{m}_{\a\b}\theta^{\b}\p
x_{m}-\frac{1}{4\ap}\g^{m}_{\a\b}\g_{m\,\g\delta}\theta^{\b}\theta^{\g}\p\theta^{\delta},\quad
\Pi^{m}=\p x^{m}+\frac{1}{2}\theta\g^{m}\p\theta,\\
\\
&& d_{\a}(z)d_{\b}(w)\sim-\frac{2}{\ap}\frac{\g^{m}_{\a\b}\Pi_{m}}{z-w},\quad
d_{\a}(z)\Pi^{m}(w)\sim \frac{\g^{m}_{\a\b}\p\theta^{\b}}{z-w},\\
\\
&&\qquad \qquad d_{\a}(z)f(\theta(w),x(w))\sim
(z-w)^{-1}D_{\a}f(\theta(w),x(w)),
\end{eqnarray*}
where
\begin{equation*}
D_{\a}=\frac{\p}{\p\theta^{\a}}+\frac{1}{2}\theta^{\b}\g^{m}_{\a\b}\p_{m},
\end{equation*}
is the covariant super-derivate on $\mathbb{R}^{10}$. The supersymmetry generator is
\begin{equation*}
q_{\a}=\int\d z\,(p_{\a}+\frac{1}{\ap}\g^{m}_{\a\b}\theta^{\b}\p
x_{m}+\frac{1}{12\ap}\g^{m}_{\a\b}\g_{m\,\g\delta}\theta^{\b}\theta^{\g}\p\theta^{\delta})
\end{equation*}
and it satisfies the algebra
\begin{equation}
\{q_{\a},q_{\b}\}=\frac{2}{\ap}\g^{m}_{\a\b}\int\d z\,\p
x_{m},\quad\[q_{\a},\Pi^{m}(z)\]=0,\quad \{q_{\a},d_{\b}(z)\}=0.
\end{equation}

The construction of the $b$-ghost is such that \cite{nathan topological}\cite{nathan nikita
multiloop}
\begin{equation*}
\{Q,b(z)\}=T(z),
\end{equation*}
where
\begin{equation*}
Q=\int\d z\,(\lambda^{\a}d_{\a}+\bar\omega^{\a}r_{\a}),\quad
T(z)=-\frac{1}{\ap}\p x^{m}\p
x_{m}-p_{\a}\p\theta^{\a}+\omega_{\a}\p\lambda^{\a}+\bar\omega^{\a}\p\bar\lambda_{\a}-s^{\a}\p
r_{\a}.
\end{equation*}
Since $Q$ and $T$ are space time dimensionless so is $b$, which is given
by
\begin{eqnarray*}
b&=&
s^{\a}\p\bar\lambda_{\a}+\frac{\bar\lambda_{\a}(2\Pi^{m}(\g_{m}d)^{\a}-N_{mn}(\g^{mn}\p\theta)^{\a}-
J_{\lambda}\p\theta^{\a}-\frac{1}{4}\p^{2}\theta^{\a})}{4(\lambda\bar\lambda)}\\
&&\,\,+\frac{(\bar\lambda\g^{mnp}r)(\frac{\ap}{2}d\g_{mnp}d+24N_{mn}\Pi_{p})}{
192(\lambda\bar\lambda)^{2}}
-\frac{\frac{\ap}{2}(r\g_{mnp}r)(\bar\lambda\g^{m}d)N^{np}}{16(\lambda\bar\lambda)^{3}}\\
&&\,\,+\frac{\frac{\ap}{2}(r\g_{mnp}r)(\bar\lambda\g^{pqr}r)N^{mn}N_{qr}}{128(\lambda\bar\lambda)^{4
}}.
\end{eqnarray*}
In order to build the vertex operators we use the following $\mathcal{N}=1$ SYM $\theta$
expansions \cite{mafra tesis}\cite{nathan multiloop}\cite{expansion}
\begin{eqnarray}
A_{\a}(x,\theta)&=&\frac{1}{2}a_{m}(\g^{m}\theta)_{\a}-\frac{1}{3}(\xi\g_{m}\theta)(\g^{m}\theta)_{
\a}
-\frac{1}{32}F_{mn}(\g_{p}\theta)_{\a}(\theta\g^{mnp}\theta)+...\\
A_{m}(x,\theta)&=&a_{m}-(\xi\g_{m}\theta)-\frac{1}{8}(\theta\g_{m}\g^{pq}\theta)F_{pq}
+\frac{1}{12}(\theta\g_{m}\g^{pq}\theta)(\p_{p}\xi\g_{q}\theta)+...\\
W^{\a}(x,\theta)&=&\xi^{\a}-\frac{1}{4}(\g^{mn}\theta)^{\a}F_{mn}+\frac{1}{4}(\g^{mn}\theta)^{\a}
(\p_{m}\xi\g_{n}\theta)+...\\
\mathcal{F}_{mn}(x,\theta)&=&F_{mn}-2(\p_{[m}\xi\g_{n]}\theta)+\frac{1}{4}(\theta\g_{[m}\g^{pq}
\theta)\p_{n]}F_{pq}+...\,\,\, .
\end{eqnarray}
Here $\xi^{\a}(x)=(2/\ap)^{1/2}\chi^{\a}e^{ik\cdot x}$, where
$[\chi^{\a}]=1/2$ and $a_{m}=e_{m}e^{ik\cdot x}$, where $[e_{m}]=0$.
$F_{mn}=2\p_{[m}a_{n]}$ is the curvature and $[F_{mn}]=-1$. The dimensions of the
superfields are
\begin{equation*}
[A_{\a}]=1/2,\quad [A_{m}]=0,\quad [W^{\a}]=-1/2,\quad
[\mathcal{F}_{mn}]=-1,
\end{equation*}
hence the massless vertex operators have the following dimensions
\begin{equation}
[V]=[\lambda^{\a}A_{\a}]=1,\,\,
[U]=[\p\theta^{\a}A_{\a}+A_{m}\Pi^{m}+\frac{\ap}{2}d_{\a}W^{\a}+\frac{\ap}{4}N_{mn}\mathcal{F}_{mn}]
=1\,\, ,
\end{equation}
where $U$ satisfies $QU=\p(\lambda^{\a}A_{\a})$. These vertex
operators have the same normalization as the vertex operators of \cite{dhoker phong s
duality}, therefore we can compare the amplitudes in a straight forward way. For
example, the closed superstring massless operator in the NS-NS sector
is \cite{dhoker phong s duality}
\begin{equation}
V=e_{m}\bar e_{n}\int\d^{2}z\,(\p
x^{m}+ik\cdot \psi_{+}\psi^{m}_{+})(\bar\p x^{n}+ik\cdot
\psi_{-}\psi^{n}_{-})e^{ik\cdot x}\,\, ,
\end{equation}
where the dimension of $V$ is two if the dimension of the
polarization vectors is zero.

\section{Four point 1-loop massless amplitude}

Using the normalization of the previous Section we will compute the one loop amplitude for
4-massless
vertex operator in the NS-NS sector. Although the general structure of this Section can be found in
the references \cite{polchinski vol 1}\cite{nathan multiloop}\cite{mafra 1 loop}\cite{mafra tesis},
we include it to justify the normalization of the measures and to find the overall constant factor
for the
amplitude, which has not been computed.

As non-minimal pure spinor formalism is a critical topological string, then one can use the
bosonic
string prescription for computing scattering amplitudes \cite{nathan topological}\cite{vafa}. So the
four points
1-loop massless amplitude is given by
\begin{equation}
\mathcal{A}=\frac{1}{2}\k^{4}\int_{\mathcal{M}_{1}}\d^{2}\tau\,\,\, \langle\Big| \mathcal{N}\,\,
(b,\mu)\,\, V(z_{1})\,\, \prod_{i=2}^{4}\int\d z_{i}\,U(z_{i})\Big|^{2}\rangle\, ,
\end{equation}
where $\mathcal{M}_{1}=\text{H}/PSL(2,\mathbb{Z})$ is the fundamental region, $\mu$ is the Beltrami
differential, $\mathcal{N}$ is a regulator, $z_{1}$ is a fixed point and finally, $\k$ is the
normalization constant of the massless vertex operator. Its precise  value will not be needed here.
The $1/2$ factor is needed
because the total group of automorphism on the torus is $SL(2,\mathbb{Z})$ instead of
$PSL(2,\mathbb{Z})$ \cite{polchinski 1 loop bosonic}\cite{polchinski vol 1}. As the amplitude is
computed
using the bosonic string prescription, we
must take in account the normalization of the inner product between the $b$-ghost and the Beltrami
differential
in the same way as in bosonic string theory \cite{polchinski vol 1}
\begin{equation}
(b,\mu)=\frac{1}{2\pi}\int \d^{2}z\,\,b(z)\mu_{\bar z}^{z}=\frac{1}{\pi}b(0),
\end{equation}
where $\mu_{\bar z\bar z}=1/2\tau_{2}$.

\subsection{Review of the $x^{m}(z,\bar z)$ fields contribution} 

In this Subsection we compute the $x^{m}(z,\bar z)$ contribution and justify, in a natural
way, the normalization of the integration measures.

In order to compute the $x^{m}(z,\bar z)$  contribution we expand
it in terms of a complete set $X_{I}(z,\bar z)$ of eigenfunctions of
the worldsheet Laplacian operator                                                                 
\begin{eqnarray*}
x^{m}(z,\bar z)&=&\sum_{I}x_{I}^{m}X_{I}(z,\bar z),\\
\p\bar\p X_{I}(z,\bar z)&=&-\lambda_{I}^{2}X_{I}(z,\bar z)\\
\int_{\Sigma_{g}}\d^{2}z\,X_{I}(z,\bar z)X_{J}(z,\bar
z)&=&\delta_{IJ}.
\end{eqnarray*}
The bosonic contribution is given by \cite{polchinski vol 1}

\begin{eqnarray}\label{bc}
\<\prod_{i=1}^{4}:e^{k_{i}\cdot x}:\>&=&\prod_{I m}\int\frac{\d
x^{m}_{I}}{\sqrt{2\pi^{2}\a^{\prime}}}\text{exp}\[-\frac{1}{2\pi\a^{\prime}}\sum_{I\neq
0}\(\lambda_{I}^{2}x_{I}\cdot x_{I}
-2\pi\ap ix_{I}\cdot J_{I}\)+ ix_{0}\cdot J_{0}\]\\
&=&(2\pi)^{10}\delta^{(10)}(J_{0})\(2\pi^{2}\a^{\prime}\text{det}^{\prime}\p\bar\p\)^{-5}
\text{exp}\[-\sum_{I\neq
0}\frac{\pi\a^{\prime}}{2\lambda_{I}^{2}}J_{I}\cdot J_{I}\]
\end{eqnarray}
\\
where

\begin{eqnarray}
J^{m}(z,\bar z)&=&\sum_{i=1}^{4}k_{i}^{m}\delta^{(2)}(z,\bar z)=\sum_{I}J_{I}^{m}X_{I}(z,\bar z)\\
J_{I}^{m}&=&\int_{\Sigma_{g}}\d^{2}z\,J^{m}(z,\bar z)X_{I}(z,\bar
z).
\end{eqnarray}
In particular
\begin{equation*}
J_{0}^{m}=X_{0}\int_{\Sigma_{g}}\d^{2}z\,J^{m}(z,\bar
z)=X_{0}\sum_{i=1}^{4}k_{i}^{m},
\end{equation*}
thus, we have

\begin{equation*}
\<\prod_{i=1}^{4}:e^{k_{i}\cdot
x}:\>=(2\pi)^{10}\delta^{(10)}(X_{0}k)\(2\pi^{2}\a^{\prime}\text{det}^{\prime}\p\bar\p\)^{-5}
\text{exp}\[-\frac{1}{2}\sum_{i\neq j}k_{i}\cdot k_{j}\sum_{I\neq
0}\frac{\pi\a^{\prime}}{\lambda_{I}^{2}}X_{I}(z_{i},\bar
z_{i})X_{I}(z_{j},\bar z_{j})\]
\end{equation*}
\\
where $k=\sum_{i=1}^{4}k_{i}^{m}$. The term
\begin{equation*}
\sum_{I\neq 0}\frac{\pi\a^{\prime}}{\lambda_{I}^{2}}X_{I}(z_{i},\bar
z_{i})X_{I}(z_{j},\bar z_{j})
\end{equation*}
is the Green's function and it satisfies the differential equation
\begin{eqnarray}
-\frac{1}{\pi\a^{\prime}}\p\bar\p G(z,w)&=&\sum_{I\neq
0}X_{I}(z,\bar
z)X_{I}(w,\bar w)\\
&=&\delta^{(2)}(z-w)-X_{0}^{2}.
\end{eqnarray}
In the torus we have defined the normalization of the $X_{0}$ mode
to be
\begin{equation}
X_{0}^{2}=(2\tau_{2})^{-1},
\end{equation}
such that
\begin{equation}
||X_{0}||^{2}=X_{0}^{2}\int_{\Sigma_{g}}\d^{2} z\,=1
\end{equation}
where $\int_{\Sigma_{g}}\d^{2}z=2\tau_{2}$.
\\
With this normalization, the Green's function for the torus is
given by\cite{dhoker phong perturbation}

\begin{eqnarray*}
G(z,w,\tau)&=&-\frac{\a^{\prime}}{2}\text{ln}|E(z,w)|^{2}+\frac{\a^{\prime}\pi}{4\tau_{2}}(z-\bar
z-w+\bar w)^{2}\\
&=&-\frac{\a^{\prime}}{2}\text{ln}|E(z,w)|^{2}+\frac{2\a^{\prime}\pi}{\tau_{2}}\text{Im}z\,\text{Im}
w,
\end{eqnarray*}
and therefore the final expression for the bosonic contribution is \cite{dhoker phong
perturbation}\cite{dhoker phong s duality}

\begin{equation*}
\<\prod_{i=1}^{4}:e^{k_{i}\cdot
x}:\>=(2\pi)^{10}(2\tau_{2})^{5}\delta^{(10)}(k)\(2\pi^{2}\a^{\prime}\text{det}^{\prime}\p\bar\p\)^{
-5}
\prod_{i<j}|E(z_{i},z_{j})|^{\a^{\prime}k_{i}\cdot k_{j}}
\text{exp}\[-k_{i}\cdot
k_{j}\frac{2\pi\a^{\prime}}{\tau_{2}}\text{Im}z_{i}\,\text{Im}z_{j}\].
\end{equation*}

The factors $(2\pi^{2}\ap)^{-1/2}$ of the integration measure of (\ref{bc}) come from treating the
$x^{m}(z,\bar z)$ action in a first order formalism \cite{losev 1 order}. To see this, let's take
the action
\begin{equation}\label{ap}
S=\frac{1}{\pi\a^{\prime}}\int\d^{2}z\,(g^{i\bar j}p_{i}p_{\bar
j}+p_{i}\bar\p x^{i}+p_{\bar i}\p x^{\bar i})
\end{equation}
where the index $i,\bar i=1,...,5$, $p_{i}$ and $p_{\bar i}$ are (1,0) and
(0,1) forms with conformal weight (1,0) and (0,1) respectively and $g^{i\bar j}=\de^{i \bar j}$.

In this first order action we can easily see that the conjugate momenta of the $x^{i}$
and $x^{\bar i}$ fields are
$P_{i}:=p_{i}/\pi\ap$ and $P_{\bar i}:=p_{\bar i}/\pi\ap$
respectively, so  the Dirac brackets ($DB$) are
\begin{eqnarray*}
&&\[P_{i}(\sigma),x^{j}(\sigma^{\prime})\]_{DB}=\[\frac{p_{i}(\sigma)}{\pi\ap},x^{j}(\sigma^{\prime}
)\]_{DB} =i\delta_{i}^{j}\delta(\sigma-\sigma^{\prime}),\\
&&\[P_{\bar i}(\sigma),x^{\bar
j}(\sigma^{\prime})\]_{DB}=\[\frac{p_{\bar i}(\sigma)}{\pi\ap},x^{\bar
j}(\sigma^{\prime})\]_{DB}=i\delta_{\bar i}^{\bar
j}\delta(\sigma-\sigma^{\prime}).
\end{eqnarray*}
In quantum mechanics, because of the commutator relation $[p,x]=i$ one has the
identity
\begin{equation}
\int\frac{\d x}{\sqrt{2\pi}}\frac{\d p}{\sqrt{2\pi}}e^{-ipx}=1,
\end{equation}
and the integration measure on the phase space in the path integral is \cite{polchinski vol 1}
\begin{equation}
\frac{\d x}{\sqrt{2\pi}}\frac{\d p}{\sqrt{2\pi}}.
\end{equation}
In the same way, the measure on the phase space in the path integral
for the action ($\ref{ap}$) is
\begin{eqnarray*}
\prod_{z\bar z}\prod_{i,\bar i,j,\bar j}\frac{\d
P_{i}}{\sqrt{2\pi}}\frac{\d P_{\bar i}}{\sqrt{2\pi}} \frac{\d
x^{j}}{\sqrt{2\pi}}\frac{\d x^{\bar j}}{\sqrt{2\pi}}&=&\prod_{z\bar
z}\prod_{i,\bar i,j,\bar j}\frac{\d
p_{i}}{\pi\ap\sqrt{2\pi}}\frac{\d p_{\bar i}}{\pi\ap\sqrt{2\pi}}
\frac{\d
x^{j}}{\sqrt{2\pi}}\frac{\d x^{\bar j}}{\sqrt{2\pi}}\\
&=&\prod_{z\bar z}\prod_{i,\bar i,j,\bar j}\frac{\d
p_{i}}{\sqrt{2\pi^{2}\ap}}\frac{\d p_{\bar
i}}{\sqrt{2\pi^{2}\ap}}\frac{\d x^{j}}{\sqrt{2\pi^{2}\ap}}\frac{\d
x^{\bar j}}{\sqrt{2\pi^{2}\ap}},
\end{eqnarray*}
hence if we compute the integral by $p_{i},\,p_{\bar i}$ fields we get (\ref{bc}).

We can see that the $p^{m}$ fields have $10g$ zero modes  in a
Riemann surface of genus g and their normalizations do not  affect the answer.

Note that in this first order formalism the number of zero modes of the bosonic fields, including
the pure spinors at the right and left sectors, is equal to zero modes of the fermionic fields
\begin{eqnarray}
&&\text{bosonic}\,\,\,\,\,\,\left\{      
 \begin{matrix}
  x^{m} & p^{m} & \l^{\a} & \o_{\a} & \lb_{\a} & \ob^{\a} & \hat\l^{\a} &
\hat\o_{\a} & \hat\lb & \hat\ob^{\a} \\ 
        10    & 10g   & 11      & 11g     & 11       & 11g      & 11       & 11g  & 11 & 11g\\
 \end{matrix}\right.\\
&&\text{fermionic}\,\,\left\{      
 \begin{matrix}
  \theta^{\a} & p_{\a} & \hat\theta^{\a} & \hat p_{\a} & r_{\a} & s^{\a} & \hat r_{\a} &
\hat s^{\a} \\ 
        16    & 16g   & 16      & 16g     & 11       & 11g      & 11       & 11g  \\
 \end{matrix}\right. \nonumber.
\end{eqnarray}

\subsection{Pure spinors and                                             
$p_{\a},\,\theta^{\a}$ fields contribution}


First, we will compute the contribution of the non zero modes to the amplitude and we will show
explicity that in the non-minimal formalism it is not necessary to compute functional determinants.

The action of the pure spinors in a chart is given by \cite{nathan covariant superstring}
\begin{equation*}
S=-\frac{1}{2\pi}\int_{\Sigma_{g}}\d^{2}z\,(\beta\bar\p                                   
\gamma+\frac{1}{2}v^{ab}\bar\p u_{ab}+\bar\beta\bar\p
\bar\gamma+\frac{1}{2}\bar v_{ab}\bar\p \bar u^{ab}).
\end{equation*}
When the Riemann surface $\Sigma_{g}$ is the torus, all the elements of the set
$\{\g,\bar\g,\beta,\bar\b,
u_{ab},\bar u_{ab},v^{ab},\bar v^{ab}\}$ have
one zero mode only \cite{dhoker phong perturbation}.

The contribution of the non-zero modes is given by
\begin{eqnarray}\label{cnzm}
&&\prod_{I\neq
0}\int\frac{[\d\beta_{I}]}{\sqrt{4\pi^{2}}}\wedge\frac{[\d\gamma_{I}]}{\sqrt{4\pi^{2}}}
\bigwedge_{a<b,\,c<d}\frac { [ \d
v^{ab}_{I}]}{\sqrt{4\pi^{2}}}\frac{[\d
u_{cd\,I}]}{\sqrt{4\pi^{2}}}\wedge
\frac{[\d\bar\beta_{I}]}{\sqrt{4\pi^{2}}}\wedge\frac{[\d\bar\gamma_{I}]}{\sqrt{4\pi^{2}}}
\bigwedge_{e<f,\,g<h}\frac { [ \d
\bar v_{ef\,I}]}{\sqrt{4\pi^{2}}}\frac{[\d
\bar u^{gh}_{I}]}{\sqrt{4\pi^{2}}}\nonumber\\
&&\text{ exp}\(-\frac{1}{2\pi}\sum_{I\neq
0}\lambda_{I}(\beta_{I}
\gamma_{I}+\frac{1}{2}v^{ab}_{I} u_{ab\,I}+\bar\beta_{I}
\bar\gamma_{I}+\frac{1}{2}\bar v_{ab\,I} \bar u^{ab}_{I} )\)
\end{eqnarray}
where the $\{\lambda_{I}\}$ are the eigenvalues of the $\bar\p$
operator and we write the measure in the same way as in the previous
Section. We can write the argument of the exponential function in the following form (for example
for
the ($\g,\b$) fields)
\begin{equation}
\text{ exp}\(-\frac{1}{2\pi}\sum_{I\neq
0}\lambda_{I}(\beta_{I}
\gamma_{I}+\bar\beta_{I}
\bar\gamma_{I} )\)=\text{
exp}\(\frac{1}{2\pi}V^{\dag}\,M\,V\)\,\, , 
\end{equation}
where  $V^{T}=(\g_{I},\bar\beta_{I})$, $M$ is the matrix
\begin{equation}
M:=\(
 \begin{matrix}
0 & A    \\
A & 0
 \end{matrix}
\)    \,\, ,
\end{equation}
and $A$ is the matrix $A:=\text{diag}(\lambda_{I})$. The same happens for the ($v^{ab},u_{cd}$)
fields.
Therefore the non-zero modes contribution of the pure spinors is $(\text{det }\pb)^{-22}$.

Although we computed the path integral of the pure spinors in a particular chart and gauge, the
answer is
correct because the $\{\g=0\}=SO(10)/U(5)$ space has measure zero with respect to the pure
spinors space.

The integration measure for the $I^{\text{th}}$ mode can be written in a covariant way \cite{mafra 5
points}\cite{mafra tesis} as follows

\begin{eqnarray}\label{lom}
&&[\d \o^{I}]=\frac{(4\pi{^2})^{-11/2}}{11!5!}(\l_{I} \ga{m})_{\a_{1}}(\l_{I}
\ga{n})_{\a_{2}}(\l_{I}
\ga{p})_{\a_{3}}(\g_{mnp})_{\a_{4}\a_{5}}\epsilon^{\a_{1}...\a_{5}\delta_{1}...\delta_{11}}
\d\o_{\de_{1}}^{I}\wedge...\wedge\d\o_{\de_{11}}^{I}\\
&&[\d \l_{I}](\l_{I} \ga{m})_{\a_{1}}(\l_{I} \ga{n})_{\a_{2}}(\l_{I}
\ga{p})_{\a_{3}}(\g_{mnp})_{\a_{4}\a_{5}}=\frac{(4\pi{^2})^{-11/2}}{11!}\epsilon_{\a_{1}...\a_{5}\r_
{1}...\r_{11}}
\d\l^{\r_{1}}_{I}\wedge...\wedge\d\l^{\r_{11}}_{I},\nonumber
\end{eqnarray}
from which we can easily see that $[\d\l_{I}]$ and $[\d\o^{I}]$ have ghost number 8 and -8,
respectively.
Taking the wedge product we get
\begin{equation}
[\d\l_{I}]\wedge[\d\o^{I}]=\frac{(4\pi{^2})^{-11}}{11!}\d\l^{\a_{1}}_{I}\wedge\d\o_{\a_{1}}^{I}
\wedge...\wedge\d\l^{\a_{11} }_{I} \wedge\d\o_{\a_{11}}^{I}.
\end{equation}
In the chart $U_{+++++}=\{\l^{+}\neq 0\}$
\begin{eqnarray*}
\l^{\a}&=&(\l^{+},\l^{ab},\l^{a})\, ,\\
\l^{+}&=&\g\,,\quad \l_{ab}=\g u_{ab}\,,\quad \l^{a}=-\frac{1}{8}\g
\epsilon^{abcde}u_{bc}u_{de}\\
\end{eqnarray*}
and in the gauge $\o_{a}=0$
\begin{eqnarray*}
\o_{\a}&=&(\o_{+},\o^{ab},\o_{a})\qquad a,\,b=1,\,2,...,5\\
\o_{+}&=&\b-\frac{1}{2\g}v^{ab}u_{ab}\, ,\quad\o^{ab}=\frac{1}{\g}v^{ab}\,,\quad \o_{a}=0,\\ 
\end{eqnarray*}
we have the measure in the
form desired
\begin{equation}
[\d\l_{I}]\wedge[\d\o^{I}]=(4\pi{^2})^{-11}\bigwedge_{a<b,\,c<d}\d\g_{I}\d u_{ab}^{I}\d\b_{I}\d
v_{I}^{cd}.
\end{equation}

For the $\lb_{\a}$ and $\bar\o^{\a}$ fields we
define the measures $[\d\lb^{I}]$ and
$[\d\ob_{I}]$ for the $I^{\text{th}}$ mode in the following form
\begin{eqnarray}\label{lobm}
&&[\d \ob_{I}](\l_{I} \ga{m})_{\a_{1}}(\l_{I} \ga{n})_{\a_{2}}(\l_{I}
\ga{p})_{\a_{3}}(\g_{mnp})_{\a_{4}\a_{5}}=(4\pi^{2})^{-11/2}\frac{(\l_{I}\lb^{I})^{3}}{11!}\epsilon_
{ \a_ { 1 } ...\a_ { 5 } \delta_ { 1 } ...\delta_{11}}
\d\ob^{\de_{1}}_{I}\wedge...\wedge\d\ob^{\de_{11}}_{I}\nonumber
\\
&&[\d \lb^{I}]=\frac{(4\pi^{2})^{-11/2}}{11!5!(\l_{I}\lb^{I})^{3}}(\l_{I} \ga{m})_{\a_{1}}(\l_{I}
\ga{n})_{\a_{2}}(\l_{I}
\ga{p})_{\a_{3}}(\g_{mnp})_{\a_{4}\a_{5}}\epsilon^{\a_{1}...\a_{5}\r_{1}...\r_{11}}
\d\lb_{\r_{1}}^{I}\wedge...\wedge\d\lb_{\r_{11}}^{I}, 
\end{eqnarray}
so
\begin{equation}
[\d\lb^{I}]\wedge[\d\ob_{I}]=\frac{(4\pi^{2})^{-11}}{11!}\d\lb^{I}_{\a_{1}}\wedge\d\ob_{I}^{\a_{1}}
\wedge...\wedge\d\lb_{\a_{11}}^{I} \wedge\d\ob^{\a_{11}}_{I},
\end{equation}
as expected.

The contribution of the fields of opposite worldsheet
chirality is $(\text{det}^{\prime}\p)^{-22}$. So, the contribution of the non
zero modes  of the pure spinors is
$(\text{det}^{\prime}\p\bar\p)^{-22}$.


From the action of the $p_{\a},\,\theta^{\a}$ fields
\begin{equation}
S_{p\t}=\frac{1}{2\pi}\int\d^{2}z\,\,p_{\a}\pb\t^{\a}\, ,
\end{equation}
we get the anticommutation relation
\begin{equation}
\left\{P_{\a}(\sigma),\t^{\b}(\sigma^{\prime})\right\}_{DB}:=\left\{\frac{p_{\a}(\sigma)}{2\pi},\t^{
\b } (\sigma^{\prime})\right\}_{DB}=
\delta_{\a}^{\b}\de(\sigma-\sigma^{\prime}).
\end{equation}
Therefore, the measure of the phase space in the path integral is \cite{polchinski vol 1}
\begin{equation}\label{ptm}
\prod_{z,\bar z}\prod_{\a\b}\d P_{\a}\,\d\t^{\b}=\prod_{z,\bar
z}\prod_{\a\b}(2\pi\d p_{\a})\,\d\t^{\b}=\prod_{z,\bar
z}\prod_{\a\b}(\sqrt{2\pi}\d p_{\a})\,(\sqrt{2\pi}\d\t^{\b}),
\end{equation}
\\
and the contribution of the non zero modes of $p_{\a}$ and $\theta^{\a}$
fields is given by
\begin{eqnarray*}
&&\prod_{\a\,\beta}\int[\d P_{\a}]^{\prime}[\d
\theta^{\beta}]^{\prime}\text{exp}\(-\frac{1}{2\pi}\int_{\Sigma_{g}}\d^{2}z\,p_{\a}\bar\p\theta^{\a}
\)\\
&&= \prod_{\a\,\beta\, I\neq 0}\int(\sqrt{2\pi}\d
p_{\a\,I})(\sqrt{2\pi}\d
\theta^{\beta}_{I})\text{exp}\(-\frac{1}{2\pi}\sum_{I\neq
0}\lambda_{I}p_{\a\,I}\theta^{\a}_{I}\)\\
&&=\[\text{det}^{\prime}\(\bar\p\)\]^{16},
\end{eqnarray*}
where $p_{a\,I},\,\theta^{\a}_{I}$ are Grassmann numbers. As in the previous case, the
contribution of the fields of opposite worldsheet chirality
is $(\text{det}^{\prime} \p)^{16}$. Thus the total contribution of the fermions $p_{\a}$ and
$\theta^{\b}$  is
\begin{equation*}
[\text{det}^{\prime}(\p\bar\p)]^{16}.
\end{equation*}

For the $r_{\a}$ and $s^{\a}$ Grassmann fields we can define the covariant measure in the path
integral for the $I^{\text{th}}$ mode as \cite{mafra tesis}
\begin{eqnarray}\label{rsm}
&&[\d r^{I}]=\frac{(2\pi)^{11/2}}{11!5!}(\lb^{I} \ga{m})^{\a_{1}}(\lb^{I}
\ga{n})^{\a_{2}}(\lb^{I}
\ga{p})^{\a_{3}}(\g_{mnp})^{\a_{4}\a_{5}}\epsilon_{\a_{1}...\a_{5}\delta_{1}...\delta_{11}}
\p_{r^{I}}^{\de_{1}}...\p_{r^{I}}^{\de_{11}}
\\
&&[\d s_{I}](\lb^{I} \ga{m})^{\a_{1}}(\lb^{I} \ga{n})^{\a_{2}}(\lb^{I}
\ga{p})^{\a_{3}}(\g_{mnp})^{\a_{4}\a_{5}}=\frac{(2\pi)^{11/2}}{11!}\epsilon^{\a_{1}...\a_{5}\r_{1}
...\r_{11}}
\p^{s_{I}}_{\r_{1}}...\p^{s_{I}}_{\r_{11}}\nonumber
\end{eqnarray}
so
\begin{equation}
[\d r^{I}][\d s_{I}]=(2\pi)^{11}\p_{r^{I}}^{1}\p^{s_{I}}_{1}...\p_{r^{I}}^{11}\p^{s_{I}}_{11}.
\end{equation}
In an analogous way as the previous case we get the contribution from the non-zero modes
\begin{equation}
\(\text{det}^{\prime}\p\pb\)^{11}. 
\end{equation}
\\
Finally, the total contribution of the non-zero modes of the
$(\l^{\a},\,\o_{\a},\,\lb_{\b},\,\ob^{\b},\,r_{\b},\,s^{\b})$ fields is
\begin{equation}
\(\text{det}^{\prime}\p\pb\)^{-11}\(\text{det}^{\prime}\p\pb\)^{-11}\(\text{det}^{\prime}\p\pb\)^{16
}\(\text{det}^{\prime}\p\pb\)^{11}=\(\text{det}^{\prime}\p\pb\)^{5}. 
\end{equation}

\subsubsection{Modular invariance}

Before to compute the zero mode contribution we discuss briefly the modular invariance.
This subject is important because the zero modes normalization of
the vertex operators and the $b$-ghost contains modular parameters. 

With all the contributions that we have computed up to now, our  4-points
1-loop amplitude has the form
\begin{eqnarray}\label{amp}
&\mathcal{A}=
\frac{(2\pi)^{10}\delta^{(10)}(k)\kappa^{4}}{2\pi^{2}
\(2\pi^{2}\a^{\prime}\)^{5}}\int_{\mathcal{M}_{1}}\d^{2}\tau\,(2\tau_{2})^{5}\prod_{k=1}^{3}\int\d^{
2}z_{k}
\prod_{i<j=1}^{4}|E(z_{i},z_{j})|^{\a^{\prime}k_{i}\cdot k_{j}}
\text{exp}\[-k_{i}\cdot
k_{j}\frac{2\pi\a^{\prime}}{\tau_{2}}\text{Im}z_{i}\,\text{Im}z_{j}\]\nonumber\\
&\Bigg|\(\frac{\ap}{2}\)^{4}\int[\d r][\d s][\d d][\d \t][\d \l][\d\lb][\d
\o][\d\ob]e^{(-\l\lb-\ob\o-r\t+sd)}
\frac{(d\ga{pqr}d)}{192(\l\lb)^{2}}(\lb\g_{pqr}r)(\l
A_{1}dW_{2}dW_{3}dW_{4})\Bigg|_{0}^{2}
\end{eqnarray}
where the subindex ``$0$'' means that only the zero modes will be computed. 
\\
Is clear that (\ref{amp}) is not modular invariant since the scattering amplitude needs a
$(\tau_{2})^{-5}$ factor
instead of the $(\tau_{2})^{5}$ factor. The reason for this is that we have not
introduced yet
the zero modes normalization of the vertex operators (so as in the $x^{m}(z,\bar z)$ fields case).
We will show that by introducing it we get the $(\tau_{2})^{-5}$ factor and the scattering amplitude
will be
modular invariant.

On the torus all the fields have one zero mode, so we can do the
following expansion on a complete set of eigenfunctions of the
world-sheet operators $\pb$ and $\p$
\begin{eqnarray*}
\t^{\a}(z,\bar
z)&=&\t^{\a}_{0}\Lambda_{0}+\sum_{I\neq0}\t^{\a}_{I}\Lambda_{I}(z,\bar
z),\quad p_{\a}(z,\bar
z)=p_{\a}^{0}\Omega_{0}+\sum_{I\neq0}p_{\a}^{I}\Omega_{I}(z,\bar
z)\\
\l^{\a}(z,\bar
z)&=&\l^{\a}_{0}\Lambda_{0}+\sum_{I\neq0}\l^{\a}_{I}\Lambda_{I}(z,\bar
z),\quad \lb_{\a}(z,\bar
z)=\lb_{\a}^{0}\Lambda_{0}+\sum_{I\neq0}\lb_{\a}^{I}\Lambda_{I}(z,\bar
z)\\
\ob^{\a}(z,\bar
z)&=&\ob^{\a}_{0}\Omega_{0}+\sum_{I\neq0}\ob^{\a}_{I}\Omega_{I}(z,\bar
z),\quad \o_{\a}(z,\bar
z)=\o_{\a}^{0}\Omega_{0}+\sum_{I\neq0}\o_{\a}^{I}\Omega_{I}(z,\bar
z),\\
r_{\a}(z,\bar
z)&=&r_{\a}^{0}\Lambda_{0}+\sum_{I\neq0}r_{\a}^{I}\Lambda_{I}(z,\bar
z),\quad s^{\a}(z,\bar
z)=s^{\a}_{0}\Omega_{0}+\sum_{I\neq0}s^{\a}_{I}\Omega_{I}(z,\bar z),
\end{eqnarray*}
where
\begin{eqnarray*}
\int\d^{2}z\Omega_{I}(z,\bar z)\bar\Omega_{J}(\bar z,
z)&=&\delta_{IJ}\\
\int\d^{2}z\Lambda_{I}(z,\bar z)\bar\Lambda_{J}(\bar z,
z)&=&\delta_{IJ},
\end{eqnarray*}
in particular
$||\Lambda_{0}||^{2}=||\Omega_{0}||^{2}=(2\tau_{2})^{-1}$. From the
previous Section we know that only the term
$\frac{(\bar\lambda\g^{mnp}r)(\frac{\ap}{2}d\g_{mnp}d)}{192(\lambda\bar\lambda)^{2}}$
of the b-ghost can saturate the $d_{\a}$ zero modes. Since our
interests are the zero modes then we write this term as
\begin{equation}
\frac{\ap}{2}\frac{(1/2\tau_{2})^{2}}{(1/2\tau_{2})^{2}}
\frac{(\bar\lambda^{0}\g^{mnp}r^{0})(d^{0}\g_{mnp}d^{0})}{192(\lambda_{0}\bar\lambda^{0})^{2}}=
\frac{\ap}{2}
\frac{(\bar\lambda^{0}\g^{mnp}r^{0})(d^{0}\g_{mnp}d^{0})}{192(\lambda_{0}\bar\lambda^{0})^{2}} .
\end{equation}
To saturate the 11 zero modes of $r_{\a}$ we need
$10r_{\a}$ zero modes. The regulator
\begin{equation}
e^{(-\l_{0}\lb^{0}-\ob_{0}\o^{0}-r^{0}\t_{0}+s_{0}d^{0})}
\end{equation}
supplies the $10r_{\a}$ zero modes plus $10\t^{\a}$ zero modes. The
$6\t^{\a}$ zero modes necessary to saturate the $16\t^{\a}$ zero
modes come from the vertex operator
\begin{equation}
(\l A_{1}dW_{2}dW_{3}dW_{4}),
\end{equation}
so, these $6\t^{\a}$ zero modes contribute
with a factor $(2\tau_{2})^{-3}$ and the $3d_{\a}$ and $\l^{\a}$
fields contribute with $(2\tau_{2})^{-2}$. In this way the factor in the right sector is
$(2\tau_{2})^{-5}$. In the left sector
the analysis is the same, so the
total factor is $(2\tau_{2})^{-10}$ and the amplitude

\begin{eqnarray}
&\mathcal{A}=
\frac{(2\pi)^{10}\delta^{(10)}(k)\kappa^{4}}{2\pi^{2}
\(2\pi\)^{10}\(\a^{\prime}\)^{5}}\(\frac{\ap}{2}\)^{8}\int_{\mathcal{M}_{1}}\frac{\d^{2}\tau}{(\tau_
{2})^{5}}\prod_{k=1}^ { 3} \int\d^ {
2}z_{k}
\prod_{i<j=1}^{4}|E(z_{i},z_{j})|^{\a^{\prime}k_{i}\cdot k_{j}}
\text{exp}\[-k_{i}\cdot
k_{j}\frac{2\pi\a^{\prime}}{\tau_{2}}\text{Im}z_{i}\,\text{Im}z_{j}\]\nonumber\\
&\Bigg|\int[\d r][\d s][\d d][\d \t][\d \l][\d\lb][\d
\o][\d\ob]e^{(-\l\lb-\ob\o-r\t+sd)}
\frac{(d\ga{pqr}d)}{192(\l\lb)^{2}}(\lb\g_{pqr}r)(\l
A_{1}dW_{2}dW_{3}dW_{4})\Bigg|_{0}^{2}
\end{eqnarray}
is modular invariant.

\subsubsection{Contribution of the zero modes}

Now we are going to compute the zero mode contribution in the NS-NS sector where we use some of the
results given in
\cite{mafra tesis}\cite{mafra 5 points}. This calculation is totally algebraic and easy to follow
due to our choice of the integration measures. 

We rewrite the integration measure in the following way
\begin{eqnarray}\label{re}
&[\d\l_{0}]=(4\pi^{2})^{-11/2}[\d\l],\quad
[\d\o^{0}]=(4\pi^{2})^{-11/2}[\d\o],\qquad[\d\lb^{0}]=(4\pi^{2})^{-11/2}[\d\lb],\quad
[\d\ob_{0}]=(4\pi^{2})^{-11/2}[\d\ob],\nonumber\\
&[\d r^{0}]=(2\pi)^{11/2}[\d r],\quad [\d s_{0}]=(2\pi)^{11/2}[\d
s],\qquad[\d\theta_{0}]=(2\pi)^{16/2}[\d\theta],\quad [\d d^{0}]=(2\pi)^{16/2}[\d d],
\end{eqnarray}
where the measures $[\d\cdot]$  are defined from the previous Subsection, for example
\begin{equation}
[\d \l](\l_{0} \ga{m})_{\a_{1}}(\l_{0} \ga{n})_{\a_{2}}(\l_{0}
\ga{p})_{\a_{3}}(\g_{mnp})_{\a_{4}\a_{5}}=\frac{1}{11!}\epsilon_{\a_{1}...\a_{5}\r_
{1}...\r_{11}}
\d\l^{\r_{1}}_{0}\wedge...\wedge\d\l^{\r_{11}}_{0},
\end{equation}
and similarly for the others measures. For the rest of this paper the subindex ``$0$''
will be dropped out. In this new notation the scattering amplitude has the form
\begin{eqnarray*}
\mathcal{A}&=&
\frac{(2\pi)^{10}\delta^{(10)}(k)\kappa^{4}}{2\pi^{2}
\(2\pi\)^{10}\(\a^{\prime}\)^{5}}\(\frac{\ap}{2}\)^{8}\int_{\mathcal{M}_{1}}\d^{2}\tau\,(\tau_{2})^
{ -5 }
\prod_{k=1}^{3}\int\d^{
2}z_{k}
\prod_{i<j=1}^{4}|E(z_{i},z_{j})|^{\a^{\prime}k_{i}\cdot k_{j}}
\\
 &&\text{exp}\[-k_{i}\cdot
k_{j}\frac{2\pi\a^{\prime}}{\tau_{2}}\text{Im}z_{i}\,\text{Im}z_{j}\]\big|(2\pi)^{-17}\mathcal{K}
\big|^{2}\\
&=&\frac{(2\pi)^{10}\delta^{(10)}(k)\kappa^{4}}{2\pi^{2}
\(2\pi\)^{44}\(\a^{\prime}\)^{5}}\(\frac{\ap}{2}\)^{8}\int_{\mathcal{M}_{1}}\d^{2}\tau\,(\tau_{2})^
{ -5 }
\prod_{k=1}^{3}\int\d^{
2}z_{k}\\
&&\prod_{i<j=1}^{4}|E(z_{i},z_{j})|^{\a^{\prime}k_{i}\cdot k_{j}}
\text{exp}\[-k_{i}\cdot
k_{j}\frac{2\pi\a^{\prime}}{\tau_{2}}\text{Im}z_{i}\,\text{Im}z_{j}\]\big|\mathcal{K}\big|^{2}\\
\end{eqnarray*}
\\
where we have defined $\mathcal{K}$ to be
\begin{equation}
\mathcal{K}=\int[\d r][\d s][\d d][\d \t][\d \l][\d\lb][\d
\o][\d\ob]e^{(-\l\lb-\ob\o-r\t+sd)}
\frac{(d\ga{pqr}d)}{192(\l\lb)^{2}}(\lb\g_{pqr}D)(\l
A_{1}dW_{2}dW_{3}dW_{4}).
\end{equation}
\\
In order to compute the $\mathcal{K}$ factor let's remember that the measures
of
$r_{\a}$ and $s^{\a}$ are given by
\begin{eqnarray*}
&&[\d r]=\frac{1}{11!5!}(\lb \ga{m})^{\a_{1}}(\lb
\ga{n})^{\a_{2}}(\lb
\ga{p})^{\a_{3}}(\g_{mnp})^{\a_{4}\a_{5}}\epsilon_{\a_{1}...\a_{5}\delta_{1}...\delta_{11}}
\p_{r}^{\de_{1}}...\p_{r}^{\de_{11}}
\\
&&[\d s](\lb \ga{m})^{\a_{1}}(\lb \ga{n})^{\a_{2}}(\lb
\ga{p})^{\a_{3}}(\g_{mnp})^{\a_{4}\a_{5}}=\frac{1}{11!}\epsilon^{\a_{1}...\a_{5}\r_{1}...\r_{11}}
\p^{s}_{\r_{1}}...\p^{s}_{\r_{11}}.
\end{eqnarray*}
We rewrite the $[\d s]$ measure as
\begin{equation}
[\d s]=\frac{1}{2^{6}\cdot11!5!}\frac{1}{(\l\lb)^{3}}(\l
\ga{r})_{\a_{1}}(\l \ga{s})_{\a_{2}}(\l
\ga{q})_{\a_{3}}(\g_{rsq})_{\a_{4}\a_{5}}\epsilon^{\a_{1}...\a_{5}\r_{1}...\r_{11}}
\p^{s}_{\r_{1}}...\p^{s}_{\r_{11}}.
\end{equation}
Integrating the $r_{\a},\,s^{\a}$ and $d_{\a}$ variables in $\mathcal{K}$ we get
\begin{eqnarray*}
\mathcal{K}&=&\frac{1}{11!11!5!\cdot2^{9}\cdot3\cdot5}\int[\d
\t][\d \l][\d\lb][\d \o][\d\ob]e^{(-\l\lb-\ob\o)}(\lb
\ga{r})^{\a_{1}}(\lb \ga{s})^{\a_{2}}(\lb
\ga{t})^{\a_{3}}(\g_{rst})^{\a_{4}\a_{5}}\\
&&\,\,\,\epsilon_{\a_{1}...\a_{5}\delta_{1}...\delta_{11}}\t^{\de_{1}}...\t^{\de_{11}}
\frac{(2^{4}\cdot 11!5!3!)}{2^{6}\cdot
3(\l\lb)^{5}}(\lb\g_{mnp}D)(\l A_{1}(\l \ga{m}W_{2})(\l
\ga{n}W_{3})(\l \ga{p}W_{4})).
\end{eqnarray*}
In \cite{mafra tesis} the following identity was proven
\begin{equation}
(\lb\g_{mnp}D)((\l A_{1})(\l \ga{m}W_{2})(\l \ga{n}W_{3})(\l
\ga{p}W_{4}))=40(\l\lb)(\l A^{1})(\l \ga{m}W^{2})(\l
\ga{n}W^{3})\mathcal{F}^{4}_{mn},
\end{equation}
and the $\mathcal{K}$ factor takes the form
\begin{eqnarray*}
\mathcal{K}&=&\frac{
40}{11!11!5!\cdot2^{9}\cdot3\cdot5}\frac{(2^{4}\cdot
11!5!3!)}{2^{6}\cdot 3}\int[\d \t][\d \l][\d\lb][\d
\o][\d\ob]\frac{e^{(-\l\lb-\ob\o)}}{(\l\lb)^{5}}\\
&&\,\,(\lb \ga{r})^{\a_{1}}(\lb \ga{s})^{\a_{2}}(\lb
\ga{t})^{\a_{3}}(\g_{rst})^{\a_{4}\a_{5}}
\epsilon_{\a_{1}...\a_{5}\delta_{1}...\delta_{11}}\t^{\de_{1}}...\t^{\de_{11}}
(\l\lb)(\l A^{1})(\l \ga{m}W^{2})(\l
\ga{n}W^{3})\mathcal{F}^{4}_{mn}.
\end{eqnarray*}
Since we are interested in the NS-NS sector, we can use the following result found in \cite{mafra
tesis}
\begin{equation}\label{kfps}
\<(\l A^{1})(\l \ga{m}W^{2})(\l
\ga{n}W^{3})\mathcal{F}^{4}_{mn}\>=\frac{1}{2^{3}\cdot 2880}K_{0},
\end{equation}
where $K_{0}$ is the Kinematic factor of \cite{dhoker phong s duality}
\begin{equation}
K_{0}=(e_{1}\cdot e_{2})\[2tu(e_{3}\cdot e_{4})-4t(e_{3}\cdot k_{1})(e_{4}\cdot
k_{2})\]+\text{perm} .
\end{equation}
But as (\ref{kfps}) was computed using the normalization   $\<(\l \ga{m}\t)(\l
\ga{n}\t)(\l \ga{p}\t)(\t\g_{mnp}\t)\>=1$, we can write the following equality for NS-NS
sector
\begin{equation}
(\l A^{1})(\l \ga{m}W^{2})(\l \ga{n}W^{3})\mathcal{F}^{4}_{mn}\Bigg|_{\text{NS-NS}}=(\l
\ga{m}\t)(\l \ga{n}\t)(\l \ga{p}\t)(\t\g_{mnp}\t)\frac{K_{0}}{2^{3}\cdot
2880}.
\end{equation}
Now, we can integrate the $\t^{\a}$ variable in the $\mathcal{K}$ factor

\begin{eqnarray}\label{int}
\mathcal{K}&=&\frac{
40}{11!11!5!\cdot2^{9}\cdot3\cdot5}\frac{(2^{4}\cdot
11!5!3!)}{2^{6}\cdot 3\cdot2^{3}\cdot 2880}\int[\d \t][\d \l][\d\lb][\d
\o][\d\ob]\frac{e^{(-\l\lb-\ob\o)}}{(\l\lb)^{5}}(\l\lb)\nonumber\\
&&\,\,(\lb \ga{r})^{\a_{1}}(\lb \ga{s})^{\a_{2}}(\lb
\ga{t})^{\a_{3}}(\g_{rst})^{\a_{4}\a_{5}}
\epsilon_{\a_{1}...\a_{5}\delta_{1}...\delta_{11}}\t^{\de_{1}}...\t^{\de_{11}}
(\l \ga{m}\t)(\l \ga{n}\t)(\l \ga{p}\t)(\t\g_{mnp}\t)K_{0}\nonumber\\
&=&
\frac{5}{2^{4}\cdot3}K_{0}\int[\d
\l][\d\lb][\d \o][\d\ob]\frac{e^{(-\l\lb-\ob\o)}}{(\l\lb)}.
\end{eqnarray}



\section{Integration on Pure spinors space}

In order to get the full expression of the one loop amplitude we need to compute the integral on
the pure spinors space. It is not a trivial integral. Actually, if we try to solve it in a
straight forward way or using computational methods maybe we could not do it.  We will use some
tools of
algebraic geometry to solve it and we suggest the reader to read before the appendix for a better
understanding of this Section.

Let's remember that the measures $[\d\l]$ and $[\d\o]$ were defined in (\ref{lom}) and (\ref{re})
\begin{eqnarray*}
&&[\d \o]=\frac{1}{11!5!}(\l \ga{m})_{\a_{1}}(\l \ga{n})_{\a_{2}}(\l
\ga{p})_{\a_{3}}(\g_{mnp})_{\a_{4}\a_{5}}\epsilon^{\a_{1}...\a_{5}\delta_{1}...\delta_{11}}
\d\o_{\de_{1}}\wedge...\wedge\d\o_{\de_{11}},
\\
&&[\d \l](\l \ga{m})_{\a_{1}}(\l \ga{n})_{\a_{2}}(\l
\ga{p})_{\a_{3}}(\g_{mnp})_{\a_{4}\a_{5}}=\frac{1}{11!}\epsilon_{\a_{1}...\a_{5}\r_{1}...\r_{11}}
\d\l^{\r_{1}}\wedge...\wedge\d\l^{\r_{11}}
\end{eqnarray*}
and the measures $[\d\lb]$ and $[\d\ob]$ were defined in (\ref{lobm}) and (\ref{re})
\begin{eqnarray*}
&&[\d \ob](\l \ga{m})_{\a_{1}}(\l \ga{n})_{\a_{2}}(\l
\ga{p})_{\a_{3}}(\g_{mnp})_{\a_{4}\a_{5}}=\frac{(\l\lb)^{3}}{11!}\epsilon_{\a_{1}...\a_{5}\delta_{1}
...\delta_{11}}
\d\ob^{\de_{1}}\wedge...\wedge\d\ob^{\de_{11}},
\\
&&[\d \lb]=\frac{1}{11!5!(\l\lb)^{3}}(\l \ga{m})_{\a_{1}}(\l
\ga{n})_{\a_{2}}(\l
\ga{p})_{\a_{3}}(\g_{mnp})_{\a_{4}\a_{5}}\epsilon^{\a_{1}...\a_{5}\r_{1}...\r_{11}}
\d\lb_{\r_{1}}\wedge...\wedge\d\lb_{\r_{11}}.
\end{eqnarray*}
With these measures it follows
\begin{equation*}
[\d\o]\wedge[\d\ob]=\frac{(\l\lb)^{3}}{11!}\d\o_{\a_{1}}\wedge\d\ob^{\a_{1}}\wedge...\wedge\d\o_{\a_
{1}}\wedge\d\ob^{\a_{1}}=(\l\lb)^{3}
\d\o_{+}\wedge\d\ob^{+}\bigwedge_{a<b,\,c<d}\d\o^{ab}\d\ob_{cd},
\end{equation*}
where we have taken the gauge $\o^{a}=\ob_{a}=0$. Now the integral (\ref{int}) on the $\o$
and $\ob$ variables is trivial
\begin{eqnarray*}
\int\,[\d\o][\d\ob]\,\,e^{-\o\ob}&=&(\l\lb)^{3}\int\,
\d\o_{+}\wedge\d\ob^{+}\bigwedge_{a<b,\,c<d}\d\o^{ab}\d\ob_{cd}\,\,e^{-\o_{+}\ob^{+}-\frac{1}{2}\o^{
ab}\ob_{ab}}\\
&=&(\l\lb)^{3}(2\pi)^{11}.
\end{eqnarray*}

\subsubsection*{\small {INTEGRAL ON PURE SPINORS SPACE}}

From the above result we can write the integral (\ref{int}) in the following
form
\begin{eqnarray}\label{psi}
\int\,[\d\l][\d\lb][\d\o][\d\ob]\,\,\frac{e^{-\l\lb-\o\ob}}{\l\lb}&=&(2\pi)^{11}\int\,[\d\l][\d\lb]
\,(\l\lb)^{2}e^{-\l\lb} \nonumber\\
&=&(2\pi)^{11}\lim_{a\rightarrow 1}\frac{\p^{2}}{\p
a^{2}}\int\,[\d\l][\d\lb]\,e^{-a\l\lb}.
\end{eqnarray}
Thus the integral of our interest is simply
\begin{eqnarray}\label{ri}
\int\,[\d\l]\wedge[\d\lb]\,e^{-a\l\lb}
\end{eqnarray}
and next we will show that it is equal to
\begin{eqnarray}
\int\,[\d\l]\wedge[\d\lb]\,e^{-a\l\lb}=\frac{(2\pi)^{11}}{a^{8}\cdot 60}.
\end{eqnarray}
\\
We can easily see that the measure $[\d\l]\wedge[\d\lb]$ can be
written as
\begin{eqnarray*}
[\d\l]\wedge[\d\lb]&=&\frac{1}{11!(\l\lb)^{3}}\d\l^{\a_{1}}\wedge\d\lb_{\a_{1}}\wedge...\wedge\d\l^{
\a_{11}}\wedge\d\lb_{\a_{11}}\\
&=&\frac{1}{11!(\l\lb)^{3}}\p\pb(\l\lb)\wedge...\wedge\p\pb(\l\lb)\\
&=&\frac{\Omega^{11}}{11!}\,\, ,
\end{eqnarray*}
where

\begin{equation*}
\Omega=\frac{1}{(\l\lb)^{3/11}}\p\pb(\l\lb)
\end{equation*}
\\
is the K\"{a}hler form\footnote{easily we can see that $(\l\lb)$ is a
scalar function (global) on the pure spinors space.}
on the pure spinors
space in  $D=2n=10$ dimension. In the parametrization (\ref{para}) the integration measure on pure
spinors space is
\begin{equation}
\frac{\Omega^{11}}{11!}=\g^{7}\d\g\bigwedge_{a<b}\d
u_{ab}\wedge\bar\g^{7}\d\bar\g\bigwedge_{c<d}\d\bar u^{cd}.
\end{equation}
The K\"{a}hler form of the pure spinors space in any dimension is given by
\begin{equation}\label{pure spinors kf}
\Omega_{D=2n}=\frac{1}{(\l\lb)^{\frac{\text{dim}_{
\mathbb{C}}PS-c_{1}}{\text{dim}_{
\mathbb{C}}PS}}} \p\pb(\l\lb),
\end{equation}
\\
where $c_{1}=2n-2$ is the first Chern class of the tangent bundle over
$SO(2n)/U(n)$ \cite{nikita nathan character} and $\text{dim}_{\mathbb{C}}PS=\frac{n(n-1)}{2}+1$ is
the complex dimension of the pure spinors space.
\\
Writing (\ref{ri}) in the coordinates (\ref{para}) we get
\begin{eqnarray}
\int\,[\d\l][\d\lb]\,e^{-a\l\lb}=\int\,(\g\bar\g)^{7}\d\g\wedge\d\bar\g\bigwedge_{a<b,\,c<d}\d
u_{ab}\d\bar u^{cd}\,e^{-a\g\bar\g(1+\frac{1}{2}u_{ab}\bar
u^{ab}+\frac{1}{8^{2}}\epsilon^{abcde}\epsilon_{afghi}u_{bc}u_{de}\bar
u^{fg}\bar u^{hi})} \,\, .
\end{eqnarray}
\\
The $\g,\,\bar\g$ variables can be integrated easily
\begin{equation}
\int\,(\g\bar\g)^{7}\d\g\wedge\d\bar\g\,\,
e^{-b\g\bar\g}=-\frac{\p^{7}}{\p b^{7}}\int\,\d\g\wedge\d\bar\g
\,\,e^{-b\g\bar\g}=(2\pi)\cdot7!\cdot\frac{1}{b^{8}}\,\, ,
\end{equation}
where
\begin{equation}
b:= a(1+\frac{1}{2}u_{ab}\bar
u^{ab}+\frac{1}{8^{2}}\epsilon^{abcde}\epsilon_{afghi}u_{bc}u_{de}\bar
u^{fg}\bar u^{hi}),
\end{equation}
so (\ref{ri}) has now the form
\begin{eqnarray}
\int\,[\d\l]\wedge[\d\lb]\,e^{-a\l\lb}=\frac{(2\pi)\cdot7!}{a^{8}}\int_{SO(10)/U(5)}\,\a \,\,\, ,
\end{eqnarray}
where
\begin{equation}\label{20 form}
\a:=\frac{\bigwedge_{a<b,\,c<d}\d u_{ab}\d\bar
u^{cd}}{(1+\frac{1}{2}u_{ab}\bar
u^{ab}+\frac{1}{8^{2}}\epsilon^{abcde}\epsilon_{afghi}u_{bc}u_{de}\bar
u^{fg}\bar u^{hi})^{8}}
\end{equation}
is a global form on $SO(10)/U(5)$, therefore it belongs to the $H_{DR}^{20}(SO(10)/U(5))$ de-Rham
cohomology
group \cite{harris}\cite{raoul}. Note that the number 8 is the first Chern class of the
tangent
bundle over $SO(10)/U(5)$.

The $\a$-form can be written as
\begin{equation}
\a=\frac{\o^{10}}{10!} ,
\end{equation}
where
\begin{equation}
\o=-\p\bar\p\,\ln(\l\lb)
\end{equation}
and $\l$  and $\lb$ are projective pure spinors, in others words
\begin{equation}
\l\lb=1+\frac{1}{2}u_{ab}\bar
u^{ab}+\frac{1}{8^{2}}\epsilon^{abcde}\epsilon_{afghi}u_{bc}u_{de}\bar
u^{fg}\bar u^{hi},
\end{equation}
where $\{u_{ab}\}$ is a complex parametrization on
$SO(10)/U(5)$. The 2-form $\o$ is the K\"ahler form, so
$\ln\,(\l\lb)$ is the K\"ahler potential \cite{harris}. From the identity
\begin{equation*}
\p\pb=\frac{1}{2}\d(\p-\pb)
\end{equation*}
we can see that $\d\o=0$ is closed, therefore $SO(10)/U(5)$ is a K\"ahler
manifold.

From the algebraic geometry point of view, the projective pure spinors space in $d=2n=10$ is a
variety
(manifold) on the projective space 
$\mathbb{C}P^{15}$, then its K\"alher form is the pullback of the K\"ahler form
of 
$\mathbb{C}P^{15}$ given by \cite{harris}\cite{spin geometry}
\begin{equation}
\o=f^{\ast}\Omega, 
\end{equation}
where $\Omega$ is the Fubini-Study \cite{harris} metric of $\mathbb{C}P^{15}$ and 
\begin{equation}
f:\, SO(10)/U(5)\,\rightarrow\,\mathbb{C}P^{15}
\end{equation}
is the corresponding map. It is given locally on the chart $U=\{\l^{+}\neq 0\}$ by the following
five
holomorphic
homogeneous polynomials \cite{cartan}\cite{nathan covariant superstring}
\begin{equation}                                        
2\l^{+}\l^{a}-\frac{1}{4}\epsilon^{abcde}\l_{bc}\l_{de}=0,\,\quad a=1,...,5.
\end{equation}
As $SO(10)/U(5)$ is a closed manifold on $\mathbb{C}P^{15}$, then it belongs to the
$H_{20}(\mathbb{C}P^{15})=\mathbb{Z}$ homology group \cite{massey}, so the projective pure spinors
space is
proportional to the $\[\mathbb{C}P^{10}\]$ homology class because $\mathbb{C}P^{10}$ is the
generator 
of the $H_{20}(\mathbb{C}P^{15})$ homology group \cite{massey}\cite{intersection}.  The
proportionality
factor is called the
 `` $\deg$''  of a variety and it is a integer number since
$H_{20}(\mathbb{C}P^{15})=\mathbb{Z}$. 
The $\rm{degree}$ of projective pure spinors is given by
\begin{equation}
\text{\deg}(SO(10)/U(5))=\,^{\#}(SO(10)/U(5)\cdot\mathbb{C}P^{5}),
\end{equation}
where $^{\#}(SO(10)/U(5)\cdot\mathbb{C}P^{5})$ are the intersection numbers  between
$SO(10)/U(5)$ and $\mathbb{C}P^{5}$ inside $\mathbb{C}P^{15}$, hence the previous
integral can be
written as
\begin{equation}
\int_{SO(10)/U(5)}\frac{\o^{10}}{10!}=\,\text{\deg}(SO(10)/U(5))\,\,\int_{\mathbb{C}P^{10}}\frac{
\Omega^{
10 }} { 10 !}\Bigg|_{\mathbb{C}P^{10}}. 
\end{equation}
Remember that the pure spinors space is identified with the total space of the line bundle
$\mathcal{O}(-1)$; which is the inverse of the line bundle
$\mathcal{L}=\mathcal{O}(1)$ \cite{harris}\cite{nikita beta gamma}. The first Chern class
$c_{1}(\mathcal{L})$ of $\mathcal{L}$ is simply
the pullback
of the hyperplane class $H$ \cite{harris}\cite{raoul}
\begin{equation}
c_{1}(\mathcal{L})=f^{\ast}H\,  
\end{equation}
and the $\rm{degree}$ of the projective pure spinors space is given by
\begin{eqnarray}\label{grado}
\int_{SO(10)/U(5)}c_{1}(\mathcal{L})^{10}&=&\,\text{\deg}(SO(10)/U(5))\,\int_{\mathbb{C}P^{10}}H
^{10}\Big|_{\mathbb{C}P^{10}}\nonumber\\
&=&\,\text{\deg}(SO(10)/U(5))\,\int_{\mathbb{C}P^{10}}\frac{
c_{10}(T\mathbb{C}P^ { 10 })}{11}\nonumber\\
&=&\,\text{\deg}(SO(10)/U(5)),
\end{eqnarray}
where $\int_{\mathbb{C}P^{10}}
c_{10}(T\mathbb{C}P^ { 10 })$ is the Euler characteristic of $\mathbb{C}P^{10}$. 
We will compute this $\rm{degree}$ using the pure spinors character at zero level. 
The Riemann-Roch formula gives us an expression for the pure spinors character at level zero
\cite{nikita nathan character}
\begin{equation}\label{rr}
Z_{10}(t)=\int_{SO(10)/U(5)}\frac{1}{1-te^{-c_{1}(\mathcal{L})}}Td(T(SO(10)/U(5))), 
\end{equation}
where $Td(T(SO(10)/U(5)))$ is the Todd genus
\begin{equation}
Td(T(SO(10)/U(5)))=1+\frac{1}{2}c_{1}(T(SO(10)/U(5))+...\,\,\,. 
\end{equation}
Expanding $Z_{10}(t)$ near to $t=1$ or near to $\epsilon=1-t=0$, the most singular term is
\cite{nikita beta gamma}
\begin{equation}
\frac{1}{\epsilon^{11}}\int_{SO(10)/U(5)}c_{1}(\mathcal{L})^{10}.
\end{equation}
The pure spinors character can also be computed with the reducibility method, in this case the
result is \cite{nikita nathan character}\cite{aldo yuri nathan nikita}
\begin{equation}
Z_{10}(t)=\frac{1+5t+5t^{2}+t^{3}}{(1-t)^{11}}.
\end{equation}
Again, expanding near $\epsilon=0$ we get that the most singular term is
\begin{equation}
\frac{12}{\epsilon^{11}}.
\end{equation}
Comparing both results we conclude that the projective pure spinors $\rm{degree}$ is
\begin{equation}
\text{\deg}(SO(10)/U(5))=\int_{SO(10)/U(5)}c_{1}(\mathcal{L})^{10}=12.
\end{equation}
Therefore we have solved in a easy way the  integral (\ref{ri})

\begin{eqnarray}\label{if}
\int_{\mathcal{O}(-1)}\,[\d\l]\wedge[\d\lb]\,\,e^{-a\l\lb}&=&\frac{(2\pi)\cdot7!}{a^{8}}\int_{
SO(10)/U(5) } \frac { (f^{ \ast } \Omega)^{10}}{10!}\nonumber\\
&=&\frac{(2\pi)\cdot7!\cdot12}{a^{8}\cdot10!}\cdot
\int_{\mathbb{C}P^{10}}\Omega^{10}\big|_{\mathbb{C}P^{10}}\nonumber\\
&=&\frac{(2\pi)^{11}\cdot7!\cdot12}{a^{8}\cdot10!}\nonumber\\
&=&\frac{(2\pi)^{11}}{a^{8}\cdot 60}.
\end{eqnarray}
Actually, we can compute (\ref{ri}) for any dimension using the K\"ahler
form (\ref{pure spinors kf}) (see appendix)
\begin{center}
\begin{tabular}{| c |}
 \cline{1-1}   \\
$\int_{\mathcal{O}(-1)}\,[\d\l]\wedge[\d\lb]\,\,e^{-a\l\lb}=
\frac{(2\pi)^{c_{1}(T\mathbb{C}P^{n(n-1)/2})}}{a^{c_{1}(T\mathcal{Q}_{2n})}}\cdot
\frac{c_{1}(T\mathcal{Q}_{2n})!}{c_{1}(T\mathbb{C}P^{n(n-1)/2})!} \cdot
\frac{c_{1}(T\mathbb{C}P^{n(n-1)/2})}{c_{1}(T\mathcal{Q}_{2n})}\cdot \text{\deg}(\mathcal{Q}_{2n})
$\\
 \\
\cline{1-1}
\end{tabular}
\end{center}
where $c_{1}(T\mathcal{Q}_{2n})=2n-2$ is the first Chern class of the tangent bundle over
projective pure spinors space $\mathcal{Q}_{2n}\equiv SO(2n)/U(n)$,
$c_{1}(T\mathbb{C}P^{n(n-1)/2})=(n(n-1)+2)/2$ is the first Chern class of the tangent bundle over
projective space $\mathbb{C}P^{n(n-1)/2}$ and $\text{\deg}(\mathcal{Q}_{2n})$ is the \rm{degree}
of the
projective pure spinors space
\begin{equation}
[\mathcal{Q}_{2n}]=\text{\deg}(\mathcal{Q}_{2n})[\mathbb{C}P^{n(n-1)/2}].
\end{equation}

With this result, we finally have that the 4-points scattering amplitude  is

\begin{eqnarray}\label{final result}
\mathcal{A}&=&(2\pi)^{10}\delta^{(10)}(k)\,\,\frac{1}{2^{7}\pi^{2}
\(\a^{\prime}\)^{5}}\,\,\[\(\frac{\ap}{2}\)^{2}\k\]^{4}\,K_{0}\overline{K}_{0}
\int_{\mathcal{M}_{1}}
\frac { \d^ { 2 } \tau } { (\tau_ { 2 } )^{ 5 }}
\prod_{k=1}^{3}\int\d^{
2}z_{k}\prod_{i<j=1}^{4}|E(z_{i},z_{j})|^{\a^{\prime}k_{i}\cdot k_{j}}\nonumber\\
&&
\text{exp}\[-k_{i}\cdot
k_{j}\frac{2\pi\a^{\prime}}{\tau_{2}}\text{Im}z_{i}\,\text{Im}z_{j}\]
\end{eqnarray}
\\
This answer is in perfect agreement with the result found by D'hoker, Phong and Gutperle in
\cite{dhoker
phong
s
duality} up to a $(\ap/2)^{8}$ factor.
Is easy to see that this factor is needed in order to have
the right space-time
dimensions \cite{private comucation}. Hence the amplitude found in \cite{dhoker phong s duality} by
D'hoker,
Phong and
Gutperle missed this term.

\section*{Acknowledgments}
I am grateful to Carlos Mafra for useful conversations, correspondences and references. I
especially thank my advisor Nathan Berkovits for patient explanations. I would like to thank Eric
D'hoker, Michael Gutperle  and  D.H Phong for discussions on the RNS formalism.
I am also grateful to Oscar A. Bedoya, D.M Schmidtt and Rold\~ao da Rocha for reading the
manuscript.
This
work is
supported by
FAPESP Ph.D grant 07/54623-8.

\appendix

\section{Pure spinors in lower dimensions and partition function}

The aim of studying pure spinors in lower dimensions ($D=2n<10$) is to have a better feeling
of some algebraic properties of the pure spinors space.
At the end of the appendix we make some remarks and give a nice geometric interpretation of the
character of pure spinors. 

We know that in $D=4,6,8$ the projective pure spinors space are $\mathbb{C}P^{1}$, $\mathbb{C}P^{3}$
and a quadric variety embedded in $\mathbb{C}P^{7}$, respectively. 

$\mathbb{C}P^{1}$ and $\mathbb{C}P^{3}$ are the trivial cases because in $D=4,6$ the pure spinors
don't have any constraints and the pure spinors space is the
simple blow-up of the origin \cite{spin geometry} (the pure spinors space is the total space of
the line bundle
$\mathcal{O}(-1)$). In these cases the K\"alher form of the pure spinors space is simply
\begin{equation}\label{kfpsld}
\Omega=\p\pb(\l\lb),
\end{equation}
where we have used the general formula (\ref{pure spinors kf})

\begin{center}
\begin{tabular}{| c |}
 \cline{1-1}   \\
$\Omega_{D=2n}=(\l\lb)^{-\frac{\dim_{
\mathbb{C}}PS-c_{1}}{\dim_{
\mathbb{C}}PS}} \p\pb(\l\lb)
$\\
 \\
\cline{1-1}
\end{tabular}
\end{center}
and the notation
\begin{eqnarray}
\l\lb&=&\g\bar\g(1+z\bar z), \qquad\qquad\quad\,\,\,\,\,\; \text{for } D=4\,,\\
\l\lb&=&\g\bar\g(1+z\bar z+u\bar u+v\bar v), \quad \,\,\text{for } D=6\, ,
\end{eqnarray}
where $\{z\}$ parametrize $\mathbb{C}P^{1}$, $\{z,\,u,\,v\}$ parametrize $\mathbb{C}P^{3}$,
$\{\g\}$ is the fiber and $c_{1}$ is the first Chern class of projective pure spinors space. From
\cite{nikita nathan character} we can see that in $D=4,6$ the first Chern class of
the tangent bundle over  the
projective pure spinors space is         
\begin{eqnarray}\label{c1ps}
&&c_{1}(T\mathbb{C}P^{1})=2,\nonumber\\
&&c_{1}(T\mathbb{C}P^{3})=4
\end{eqnarray}
and it has the same  value of the complex dimension of the pure spinors space
$(\text{dim}_{\mathbb{C}}PS)$. 

The integration measures for the pure spinors space in $D=4,6$ are given by
\begin{eqnarray}
\frac{\Omega^{2}}{2!}&=&\omega\wedge\bar\omega\qquad \text{for }D=4\, ,\\
\frac{\Omega^{4}}{4!}&=&\omega\wedge\bar\omega\qquad \text{for }D=6\, ,
\end{eqnarray}
where
\begin{eqnarray}
\omega&=&\g\,\d\g\wedge\d z  \qquad       \qquad \qquad\,\,\,\,\, \text{for }D=4\\
\omega&=&\g^{3}\,\d\g\wedge\d z\wedge\d u\wedge\d v \qquad \text{for }D=6
\end{eqnarray}
are the holomorphic top forms, which agree with the ones of \cite{nikita beta gamma}. To compute
(\ref{c1ps}) is
very easy from the following exact sequence of
bundles (the Euler sequence)\cite{harris}\cite{spin geometry}
\begin{equation}
0\longrightarrow \mathbb{C}\longrightarrow H^{\oplus(n+1)} \longrightarrow T\mathbb{C}P^{n}
\longrightarrow 0\,\, ,
\end{equation}
where $\mathbb{C}$ is a trivial bundle, $H$ is the hyperplane class and $T\mathbb{C}P^{n}$ is the
tangent bundle on $\mathbb{C}P^{n}$. This sequence implies that
\begin{equation}
H^{\oplus(n+1)}=T\mathbb{C}P^{n}\oplus \mathbb{C}. 
\end{equation}
Therefore, the total Chern class of the tangent bundle on $\mathbb{C}P^{n}$ is
\begin{equation}
 c(T\mathbb{C}P^{n})=(1+H)^{n+1}
\end{equation}
where we have denoted the first Chern class of the hyperplane bundle $H$ with the same letter
$H$. Now it is clear that $c_{1}(T\mathbb{C}P^{n})=(n+1)H$ and
that $c_{n}(T\mathbb{C}P^{n})=(n+1)H^{n}$. As
the Euler characteristic of a complex manifold $M$ of complex dimension $n$ is \cite{harris}
\begin{equation}
\chi(M)=\int_{M} c_{n}(TM)\,\, ,
\end{equation}
then we have that
\begin{equation}\label{ch}
 \int_{\mathbb{C}P^{n}}H^{n}=1,
\end{equation}
which was used in (\ref{grado}) and (\ref{if}). 
Let's apply the previous results to the pure spinors space in $D=4$.
\\
We know that the integration measure on the pure spinors space in $D=4$ is
\begin{equation}
\frac{\Omega^{2}}{2!}=-\g\bar\g\,\d\g\wedge\d\bar\g\wedge\d z\wedge\d\bar z. 
\end{equation}
Let's integrate the function exp$\{-a\l\lb\}$, with $a\in\mathbb{R}^{+}$,
\begin{eqnarray}\label{ps4}
\int_{\mathcal{O}(-1)}[\d\l]\wedge[\d\lb]\,\,e^{-a\l\lb}&=&-\int_{\mathbb{C}^{2}}\g\bar\g\,
\d\g\wedge\d\bar\g\wedge\d z\wedge\d\bar z \,e^{-a\g\bar\g(1+z\bar z)}\nonumber\\
&=&\frac{\pi}{a^{2}i}\int_{\mathbb{C}}\frac{2}{(1+z\bar z)^{2}}\d z\wedge\d\bar z.
\end{eqnarray}
We can see that $g_{z\bar z}=2/(1+z\bar z)^{2}$ is the metric of $S^{2}$ with radius 1 on a chart
homeomorphic to $\mathbb{C}$. The area of a sphere with radius $R$
is $4\pi R^{2}$, so the integral (\ref{ps4}) is $4\pi^{2}/a^{2}$. Nevertheless we want to show how
to
compute the integral (\ref{ps4}) using simple topological properties of the projective pure spinors
space ($S^{2}$). Let's remember that the first Chern class of a complex manifold $\mathcal{M}$ is
given by the expression
\begin{equation}
c_{1}(T\mathcal{M})=\frac{i}{2\pi }\p\pb\ln\det(g_{i\bar j}),
\end{equation}
so, in our example we have
\begin{equation}
c_{1}(TS^{2})=\frac{2}{2\pi i}\frac{\d z\wedge\d\bar z}{(1+z\bar z)^{2}}. 
\end{equation}
Note that the number 2 on the numerator, which comes of the exponent of $(1+z\bar
z)^{2}$, is simply the first Chern class of the tangent bundle with respect to the hyperplane
bundle $H$ ($c_{1}(TS^{2})=2H$)\footnote{This is the same argument by which the number 8 is in the
20-form (\ref{20 form}).}, hence
\begin{equation}
H=\frac{1}{2\pi i}\frac{\d z\wedge\d\bar z}{(1+z\bar z)^{2}} 
\end{equation}
on the chart. Now, using (\ref{ch}) we can easily compute (\ref{ps4})
\begin{eqnarray}
\int_{\mathcal{O}(-1)}[\d\l]\wedge[\d\lb]\,\,e^{-a\l\lb}=\frac{2\pi}{a^{2}}2\pi\int_{\mathbb{C}}
\frac { \d z\wedge\d\bar z}{2\pi i (1+z\bar
z)^{2}}=\frac{4\pi^{2}}{a^{2}}\int_{\mathbb{C}P^{1}}H=\frac{4\pi^{2}}{a^{2}},
\end{eqnarray}
as expected.

We can get the same result (\ref{ch}) from the partition
function, for example, computing the partition function for $\mathcal{O}(-1)$ over 
$\mathbb{C}P^{n}$ 
in the zero level with the reducibility
method \cite{aldo yuri} we have
\begin{equation}
Z_{\mathcal{O}(-1)}(t)=\frac{1}{(1-t)^{n+1}}.
\end{equation}
Expanding around to $\epsilon=1-t=0$ the most singular term is
\begin{equation}
\frac{1}{\epsilon^{n+1}},
\end{equation}
and by comparing  with the Riemann-Roch formula (\ref{rr}) we get (\ref{ch}).

Now we discuss some aspects of intersection theory. It is clear that in $\mathbb{C}P^{n}$ we have
a set \{$\mathbb{C}P^{m}$\} with $m\leq n$ which is
embedded it.  It is  easy to see that these $\mathbb{C}P^{m}$'s intersect transversally of a
point
\cite{harris},
i.e
\begin{equation}\label{gh}
^{\#}(\mathbb{C}P^{m}\cdot \mathbb{C}P^{n-m})=1\,,\qquad m\leq n. 
\end{equation}
As the homology groups of $\mathbb{C}P^{n}$  are \cite{massey}
\begin{equation}
H_{2i}(\mathbb{C}P^{n})=\mathbb{Z}\,,\qquad i=1,2,...,n
\end{equation}
then by (\ref{gh}) we can take the homology generators to be the $[\mathbb{C}P^{i}]$ classes.  With
this, we define the $\rm{degree}$ of a closed variety $V$ of complex dimension $m$ by
\begin{equation}
\text{\deg}(V)=\,\,^{\#}(V\cdot \mathbb{C}P^{n-m}). 
\end{equation}
This is a topological number because it depends only on the homology class.

Now we compute the $\rm{degree}$ for projective pure spinors in $D=8$.The projective pure spinors
space
in
$D=8$ ($\mathcal{Q}_{8}$) is a hypersurface in $\mathbb{C}P^{7}$. It is given in terms of
homogeneous coordinates $\{\l^{+},\l_{12},\l_{13},\l_{14},\l_{23},\\ 
\l_{24},\l_{34},\l_{1234}\}$ on
$\mathbb{C}P^{7}$ as the zero locus of \cite{cartan}
\begin{equation}
\l^{+}\l_{1234}-\l_{12}\l_{34}+\l_{13}\l_{24}-\l_{23}\l_{14}=0.
\end{equation}
Since $\text{\deg}(\mathcal{Q}_{8})$ is the number of points where $\mathcal{Q}_{8}$ and
$\mathbb{C}P^{1}$ are intersected, if we take $\mathbb{C}P^{1}$ as the locus
$\l_{12}=\l_{13}=\l_{14}=\l_{23}=\l_{24}=\l_{34}=0$, the $\text{\deg}(\mathcal{Q}_{8})$ will be the
number of solutions of the homogeneous polynomial  
\begin{equation}
\l^{+}\l_{1234}=0. 
\end{equation}
The solutions of this polynomial are the points $[1,0,0,0,0,0,0,0]$ and $[0,0,0,0,0,0,0,1]$,
therefore
\begin{equation}
\text{\deg}(\mathcal{Q}_{8})=2.
\end{equation}
Using the partition function we get the same answer, i.e, the partition function for
$\mathcal{O}(-1)$ over $\mathcal{Q}_{8}$ is given by \cite{aldo yuri}
\begin{equation}
 Z_{\mathcal{Q}_{8}}(t)=\frac{1+t}{(1-t)^{7}}\, .
\end{equation}
Expanding near to $\epsilon=1-t=0$, the most singular term of $Z_{\mathcal{Q}_{8}}(t)$ is
\begin{equation}
 \frac{2}{\epsilon^{7}},
\end{equation}
so, by comparing with the Riemann-Roch formula (\ref{rr}) we get
\begin{equation}
\int_{\mathcal{Q}_{8}}c_{1}(\mathcal{L})^{6}=2.
\end{equation}
Actually this result was expected,  since $\mathcal{Q}_{8}$ is a hypersurface given by a
homogeneous polynomial of degree 2, then the first Chern class of the divisor $[\mathcal{Q}_{8}]$
is
\begin{equation}
 c_{1}([\mathcal{Q}_{8}])=2H,
\end{equation}
which is Poincar\'e dual to $\mathcal{Q}_{8}$ \cite{harris}\cite{raoul}. So
\begin{equation}
\int_{\mathcal{Q}_{8}}c_{1}(\mathcal{L})^{6}=\int_{\mathcal{Q}_{8}}(f^{*}H)^{6}=\int_{\mathbb{C}P^{7
}} H^ { 6 } \wedge c_{1}([\mathcal{Q}_{8}])=2\int_{\mathbb{C}P^{7
}} H^ { 7 } = 2.
\end{equation}
where $f:\mathcal{Q}_{8}\rightarrow \mathbb{C}P^{7}$ is  the embeding.
 
We now have a geometric interpretation to the result found in \cite{nikita nathan character}. In
\cite{nikita nathan character} it was shown that the partition function of pure spinors can be
written
as a
rational function\footnote{ we are only interested in the zero level.}
\begin{equation}\label{rf}
Z_{\mathcal{O}(-1)}(t)=\frac{P(t)}{Q(t)} \, ,
\end{equation}
where $P(t)$ and $Q(t)$ are polynomials. In $D=2n$ the $Q(t)$ polynomial has the form \cite{nikita
nathan character}\cite{aldo yuri nathan nikita}
\begin{equation}\label{Q}
Q(t)=(1-t)^{\text{dim}_{\mathbb{C}}PS}.
\end{equation}
In \cite{nikita nathan character} it was also shown that
$Z_{\mathcal{O}(-1)}(t)$ can be written as an infinite product ($ghost-ghost$)
\begin{equation}\label{pi}
Z_{\mathcal{O}(-1)}(t)=\prod_{n=1}^{\infty}(1-t^{n})^{-N_{n}}.
\end{equation}
The $N_{n}$ coefficients contain the information about the Virasoro central charge, ghost number
anomaly, etc
\begin{eqnarray}
\frac{1}{2}c_{\text{vir}}&=&\sum_{n} N_{n}\, ,\\
a_{\text{ghost}}&=&\sum_{n}nN_{n}.
\end{eqnarray}
From (\ref{rf}) and (\ref{pi}) we have
\begin{eqnarray}
\ln(-x)\sum_{n}N_{n}+\sum_{n}\ln(n)N_{n}+\frac{x}{2}\sum_{n}nN_{n}+\sum_{g=1}^{\infty}\frac{B_{2g}
} { 2g(2g)!}x^{2g}\sum_{n}n^{2g}N_{n}=-\ln P(e^{x}) + \ln Q(e^{x}),
\end{eqnarray}
where $\{B_{g}\}$ are the Bernoulli numbers. Replacing (\ref{Q}) in the previous expression we get
\begin{equation}
\ln (1-e^{x})^{\text{dim}_{\mathbb{C}}PS}=
(\text{dim}_{\mathbb{C}}PS)\ln(-x)+\frac{\text{dim}_{\mathbb{C}}PS}{2}x+\frac{\text{dim}_{\mathbb{C
}}PS}{24} x^{2}+...\,\,\,\,\,  .
\end{equation}
Without loss of generality we can suppose that
\begin{equation}
P(e^{x})=y+a\,e^{x}+b\,e^{2x}+c\,e^{3x}+...\,\,\,\, ,
\end{equation}
so
\begin{eqnarray}
-\ln P(e^{x})&=& -\ln P(1)-\p_{x}\ln P(x)|_{x=1}\, x+.....\\
&=& -\ln P(1)-\frac{\p_{x}P(x)|_{x=1}}{P(1)}x+..... \\
&=&-\ln (y+a+b+c+...)-\frac{a+2\,b+3\,c+...}{y+a+b+c+..}x+..... \\
\end{eqnarray}
and therefore we have
\begin{eqnarray}
 \frac{1}{2}c_{\text{vir}}&=&\sum_{n} N_{n}=\text{dim}_{\mathbb{C}}PS\, ,\\
a_{\text{ghost}}&=&\sum_{n}nN_{n}=\text{dim}_{\mathbb{C}}PS-2\frac{\p_{x}P(x)|_{x=1}}{P(1)}\, ,\\
\ln P(1)&=&-\sum_{n}\ln(n)N_{n}=\ln(\text{\deg} \mathcal{Q}_{2n}),\qquad
\mathcal{Q}_{2n}:=SO(2n)/U(n).
\end{eqnarray}
From the Riemann-Roch formula (\ref{rr}) and by expanding (\ref{rf}) with (\ref{Q}) near to
$\epsilon=1-t=0$ it is clear than $\text{\deg}(\mathcal{Q}_{2n})=P(1)$. 
\\
We know that $a_{\text{ghost}}$ is the first Chern class of $T\mathcal{Q}_{2n}$ and that the
$\text{\deg}(\mathcal{Q}_{2n})$ gives the homology class  
\begin{equation}
[\mathcal{Q}_{2n}]=\text{\deg}(\mathcal{Q}_{2n})[\mathbb{C}P^{n(n-1)/2}]\, ,
\end{equation}
in others words, the $\text{\deg}(\mathcal{Q}_{2n})$ gives us the Poincar\'e dual class of
$\mathcal{Q}_{2n}$. Noting that the homology class of $\mathcal{Q}_{2n}$  is an integer number times
the homology class of $\mathbb{C}P^{n(n-1)/2}$, we can interpret
$\text{dim}_{\mathbb{C}}PS=1+n(n-1)/2$  as the first Chern class of $T\mathbb{C}P^{n(n-1)/2}$.
Thus we have
\begin{eqnarray}
&&c_{1}(T\mathbb{C}P^{n(n-1)/2})=\sum_{n} N_{n}\, ,\\
&&c_{1}(T\mathcal{Q}_{2n})=\sum_{n}nN_{n}\, ,\\
&&\text{\deg} (\mathcal{Q}_{2n})=\exp(-\sum_{n}\ln(n)N_{n})=\(\prod_{n}n^{N_{n}}\)^{-1}.
\end{eqnarray}

With these geometric interpretation we get a geometric constraint on the coefficients of the $P(t)$
polynomial
\begin{equation}
\text{\deg}(\mathcal{Q}_{2n})\,\{c_{1}(T\mathbb{C}P^{n(n-1)/2})-c_{1}(T\mathcal{Q}_{2n})\}=2\,\p_{x}
P(x)|_{x=1}. 
\end{equation}
We can also rewrite the integration measure of the pure spinors space (\ref{pure spinors kf}) as 

\begin{equation}
\Omega_{D=2n}=(\l\lb)^{-\frac{c_{1}(T\mathbb{C}P^{n(n-1)/2})-c_{1}(T\mathcal{Q}_{2n})}{
c_{1}(T\mathbb{C}P^{n(n-1)/2})}}
\p\pb(\l\lb)=(\l\lb)^{\frac{-2\,\p_{x}
P(x)|_{x=1}}{
\text{\deg}(\mathcal{Q}_{2n})c_{1}(T\mathbb{C}P^{n(n-1)/2})}} \p\pb(\l\lb)\,,
\end{equation}
\\
where we interpret the term $\{c_{1}(T\mathbb{C}P^{n(n-1)/2})-c_{1}(T\mathcal{Q}_{2n})\}$ as a
topological deviation and find a relationship between the integration measure and the character of
the
pure spinors space.

\end{document}